# Carboxyl Carbon Quantum Dots: a Novel Type of Environmental-Friendly Scale Inhibitor

Jian Hao, Lingyun Li, Weiwei Zhao, Xiaqian Wu, Yangyang Xiao, Hongfeng Zhang, Na Tang* and Xiaocong Wang*

College of Chemical Engineering and Material Science, Tianjin University of Science and Technology, Tianjin 300457, China

* Corresponding authors: tjtangna@tust.edu.cn (N. Tang); wangxc@tust.edu.cn (X. Wang).

**Abstract:** Carbon quantum dots (CQDs) are promising nano-materials since it has smaller particle size, excellent biocompatibility and low toxicity. However, no one has found their high scale inhibition performance so far. In this paper, a new kind of green scale inhibitor, carboxyl carbon quantum dots (CCQDs), were synthesized through a simple method of thermal decomposition of citric acid. The as-prepared CCQDs have excellent scale inhibition performance for $CaSO_4$ and $BaSO_4$. With static test of scale inhibition at temperature of 0～80 ℃, the anti-scaling efficiency can reach 100% with low additions of CCQDs.

## 1. Introduction

Two of the greatest problems of the human society are those related to water shortage and the environment deterioration. In recent years, the demand of water has increased significantly with the development of industry and society, which intensifies the shortage of water resources [1]. In order to make full use of water, circulating utilization is considered as an effective way to alleviate the pressure on water supply. However, during the operation, scale deposition in recirculation cooling water system is one of the most serious problems. It is a general problem on heat exchangers, cooling water systems, reverse osmosis membrane surface, boilers, secondary oil recovery and utilizing water flooding techniques, flue gas desulphurization systems and desalination plants etc. [2-8]. Once scales have formed on the surface of equipment, the heat transfer efficiency of the system will decline dramatically, and it will greatly shorten the service lifetime of the apparatus [9]. To avoid formation of scales, the most common and effective method is to add small amount of scale inhibitors [3, 10].

At present, phosphorous compounds such as amino trimethylene phosphonic acid (ATMP) and 2-phosphonobutane -1,2,4-tricarboxylic acid (PBTCA) have been effectively used against scale formation in cooling water systems [11]. But their use is limited because these compounds become nutrients in the eutrophication process, which may lead to the massive development of biological species and has the potential to deteriorate the environment [12].

In order to avoid the deterioration of ecological environment, development of the phosphate-free and nitrogen-free scale inhibitors becomes the main focus of water treatment technology [13]. The phosphorus-free polycarboxylate such as polyacrylic acid (PAA), polymaleic acid (PMA) and polyepoxysuccinic acid (PESA) have attracted great interests, both in industry and in academia. But they have low calcium tolerance because of their reaction with calcium ions to form insoluble calcium-polymer salts [14]. Therefore, exploitation of novel environment friendly scale inhibitors is of vital importance.

On the other hand, carbon-based quantum dots (CQDs) are a newly developed class of carbon nano-materials that have attracted much interest and attention owing to their un-comparable and unique properties [15-17]. Carbon quantum dots are defined as small carbon nanoparticles with outstanding features such as good conductivity, high chemical stability, environmental friendliness, broadband optical absorption, low toxicity, strong photoluminescence (PL) emission and optical properties [18]. Carbon quantum dots components and structure determine their diverse characteristics. Most of the carboxyl moieties on the surface of the CQD give a great solubility in water and biocompatibility [19-22].

Since the raw material for preparing carbon dots is mostly carbon-containing organic matter, its surface contains a large amount of reactive functional groups (such as -OH, -COOH, etc.). These oxygen-containing functional groups are easily chelated with the scale-forming ions, and the carbonic acid dots of citric acid have good adsorption capacity, which can affect the crystallization process of the fouling substances by lattice distortion. It is speculated that the carbon dots derived from citric acid should have good scale inhibition and scale inhibition ability in industrial cooling water.

So, in this paper, we present a simple method to synthesize carboxyl modified carbon quantum dots (CCQDs) and use it as a novel type of green scale inhibitor for the first time. The superior properties of this new kind of scale inhibitor have been investigated. In comparison with traditional scale inhibitor, this new type of efficient scale inhibitor has many advantages, such as easy preparation, free of phosphorus and nitrogen, low-toxicity and biocompatibility. It is a real eco-friendly scale inhibitor and will find many applications in oil pipelines, industrial boilers, water cooling systems and desalination equipment etc.

## 2. Experimental

### 2.1 Materials

Citric acid, sodium hydroxide and anhydrous sodium sulfate were purchased from Pharmaceutical Group Co., LTD.. Anhydrous calcium chloride was bought from Tianjin Wind Boat Chemical Reagent Technology Co., LTD.. All chemicals are analytical grade and used as received without further purification.

### 2.2 Synthesis of carbon quantum dots

In a typical synthesis procedure, 100 g of citric acid was added in a 250 ml three-neck round-bottom flask and was heated at 180 ℃ for 30 h. It was melted to form a colorless liquid, followed by degas. The color then changed from colorless to yellow, orange, and brown. Upon cooling, an orange-brown high-viscosity liquid was obtained, which was stirred with 100 ml of ultrapure water and a NaOH aqueous solution (5 mol/L). Approximately 25 ml of NaOH aqueous solution was subsequently added to neutralize the solution to a pH of 7, resulting in an orange-brown solution of carboxylic acid groups modified CCQDs. Further purification was completed by dialysis bag. The product was separated as a yellow-orange powder by freeze-drying, and finally about 45 g of CCQDs were obtained from 100 g of initial materials.

## 2.3 Scale inhibition measurement

Scale inhibition efficiency was measured by national standard GB/T 16632-2008, the static scale inhibition test method was adopted to measure the ability of scale inhibition for CCQDs against $CaSO_4$ precipitation. The mass concentrations of $Ca^{2+}$ and $SO_4^{2-}$ were 6800 mg/L. The solution was heated to 40 ℃, 60 ℃ and 80 ℃ in a water bath and kept for 10 h, respectively. After cooling the solution to room temperature, EDTA standard solution was added to measure the $Ca^{2+}$ ion remained in the supernatant. The scale inhibition efficiency of CCQDs against $CaSO_4$ was calculated by a formula (as shown in formula (1)).

$$\text{scale inhibition efficiency (\%)} = \frac{V_2 - V_1}{V_0 - V_1} \times 100\% \quad (1)$$

where $V_0$ (ml) is the consumed volume of EDTA solution for titration the water sample neither addition of scale inhibitors nor $NaSO_4$, while $V_1$ and $V_2$ (mL) are the consumed volume of EDTA solution in the absence and in the presence of scale inhibitors after 10 h incubation at specific temperature, respectively.

The scale effect of CCQDs against barium sulfate was examined according to the national standard of People's Republic of China concerning the code for the design of industrial oil field water treatment (Q/SY 126-2005).

## 2.4 Characterization

A JEOL JEM-2100 transmission electron microscope (TEM) was used to obtain the morphology and the size of CCQDs. Ultraviolet–visible (UV–vis) spectra was measured on a UV-2550PC spectrophotometer using ultrapure water as solvent. Fluorescent signal of CCQDs was observed by an F-280A Fluoro-spectrophotometer.

The infrared spectra (IR) were recorded by an IS50 Fourier transform infrared spectrometer using KBr disks. Functional groups on the surface of CCQDs were measured by a K-Alpha X-ray photoelectron spectrometer. A 6100 X-ray powder diffractometer and a SU1510 scanning electron microscope (SEM) were utilized to investigate the mechanism of scale inhibition.

## 3. Results and discussion

With citric acid pyrolysis, the color gradually changed from colorless to brown. The essence is that the C-H bonds in citric acid molecules are gradually oxidized to carboxyl groups, and some C-C bonds break or recombine. So citric acid molecules are continuously carbonized, and carboxyl groups on the surface are continuously enriched to form carboxyl carbon quantum dots. **Figure 1** shows TEM image of the as-prepared CCQDs. The size is centered at 4-8 nm with a relatively broad size distribution. Regions of both graphitic and amorphous carbon were found, the graphitic regions demonstrate the lattice fringes corresponding to the interlayer spacing is 3.6 Å. The TEM result was confirmed by X-ray powder diffraction. The X-ray powder diffraction of CCQDs showed a broad feature centered at 2θ = 29.8°, which attributed to the lattice spacing of 3.6 Å, similar to the (200) reflection ($d_{002}$=3.4 Å [23]) of disordered graphitic-like species (**Figure 1, insert**).

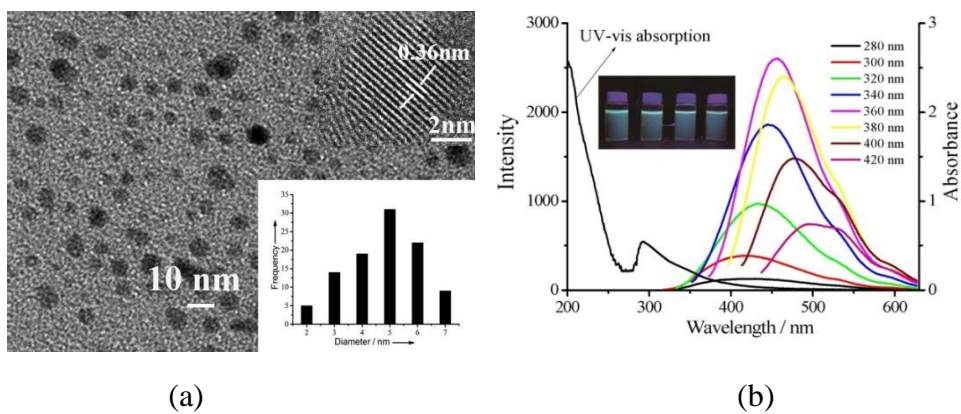

**Figure 1** (a) TEM images of CCQDs. The insert (up) is a high magnification image. The insert (down) is the size distribution of CCQDs. (b) UV-visible absorption and steady-state photoluminescence emission spectra of CCQDs in aqueous solution with varying excitation wavelengths (λ= 280, 300, 320, 340, 360, 380, 400 and 420 nm).

The optical properties of the CCQDs were explored by UV-vis and photoluminescence (PL) spectroscopy (**Figure 1b**). A broad absorption in the UV region with a tail in the near-visible region was presented in UV-vis spectra, which was assigned to various π−π* (C=C) and n−π* (C=O) transitions. Maximum emission peak was observed at λ=360 nm PL excitation wavelength. The typical excitation-wavelength-dependent behavior of PL emission was revealed. On shifting the excitation wavelength from 280 to 420 nm, the emission maximum was shifted from 407 to 493 nm. The fluorescent effect under ultraviolet lamp is blue, which related with the diameter of CCQDs. According to the fluorescent effect, it indicates that the CCQDs acquired from different pyrolysis time are about the same size.

In order to confirm the surface groups and their contents on the surface of carbon quantum dots, FTIR (**Figure 2**) and X-ray photoelectron spectroscopy (XPS) (**Figure 3**) were performed. Significant differences in FTIR spectra of citric acid and CCQDs are observed as shown in **Figure 2**. The bands at 3370 cm$^{-1}$ and 3030 cm$^{-1}$ are attributed to the stretching vibration of O-H bond and C-H bond (**Figure 2a**) [24].

There is a broad peak at 3414 cm$^{-1}$ assigned to the stretching vibration of O-H bond of hydroxyl group. Two strong features at 1398 and 1568 cm$^{-1}$ corresponding to the symmetric and anti-symmetric stretches of the carboxylate group [25], consistent with the presence of this group on the surface of the carbon dots (**Figure 2b**). By comparing the spectra of CCQDs and citric acid, it could be concluded that no residual of citric acid left in the CCQDs，the citric acid molecules had been decomposed completely.

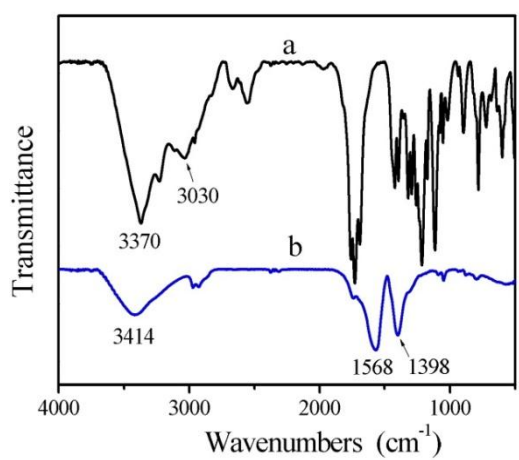

**Figure 2** FTIR spectra of citric acid (a) and CCQDs (b).

High resolution O1s spectrogram of CCQDs is shown in **Figure 3**, the intensity of O-H bond decreases with the increasing of pyrolytic time. When the pyrolytic time was 30 h, the CCQDs contained more O-C=O bonds. Combination the results of FTIR and XPS, a conclusion that there is abundant carboxylate groups on the surface of CQDs acquired from high-temperature pyrolysis of citric acid could be drawn.

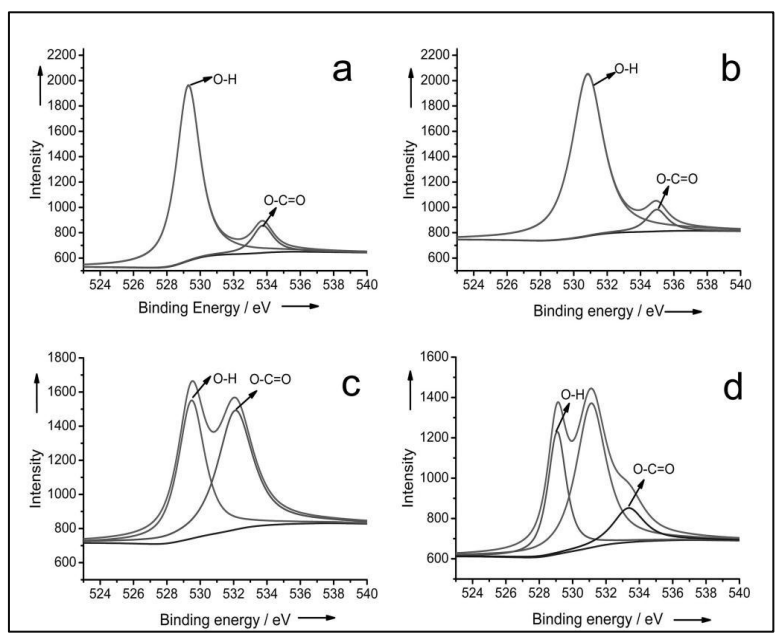

**Figure 3** High resolution O1s spectrograms of CCQDs obtained by different pyrolysis time: (a) 10 h, (b) 20 h, (c) 30 h, (d) 40 h.

Scale-inhibition performances of the CCQDs obtained from different condition were investigated by the static scale inhibition test. The scale-inhibition efficiencies of the CCQDs from different pyrolytic temperature and time have been systemically investigated. At temperature of 180 ℃, CCQDs obtained from pyrolytic time of 10 h, 20 h, 30 h and 40 h have been compared. With a result, CCQDs obtained from pyrolytic time of 30 h have the best scale-inhibition efficiency (**Figure 4a**), so it was selected to study the influence of other factors for scale inhibition.

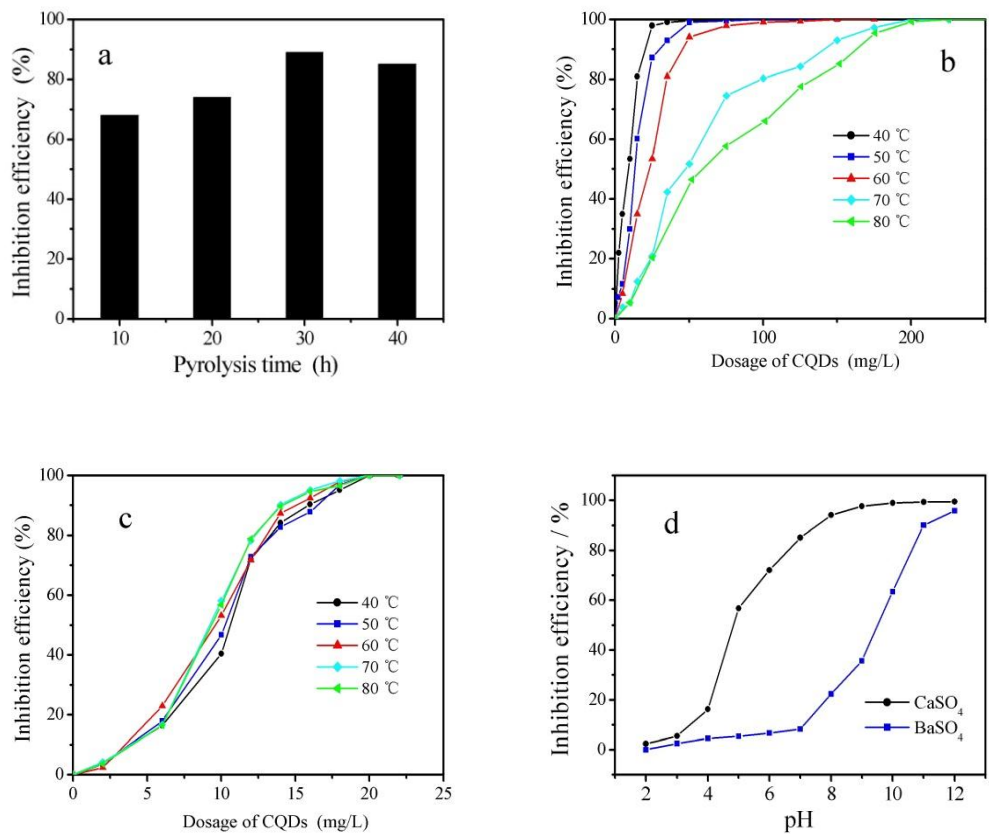

**Figure 4** The scale inhibition efficiency of CCQDs: (a) with different pyrolysis times; (b) at different temperature, $CaSO_4$; (c) at different temperature, $BaSO_4$; (d) at different pH.

Heating temperature is an important influencing factor for the scale-inhibition effect. Temperature of 40 ℃, 50 ℃, 60 ℃, 70 ℃ and 80 ℃ have been investigated. The results of calcium scales are shown in **Figure 4b**. It could be seen that the scale inhibition efficiency increases with the increasing dosage of CCQDs. For $CaSO_4$ scale, when the dosage of CCQDs is 25 mg/L, the scale inhibition efficiency reaches over 95 % at 40 ℃. To achieve a same scale-inhibition effect, higher concentration of CCQDs was needed at a higher temperature. The scale inhibition efficiency achieved 99 % at 80 ℃ with a dosage of 200 mg/L. Nevertheless, at any temperature from 40 ℃ to 80 ℃, sufficient efficiency can near or reach 100%. That can fully meet the needs in reverse osmosis membranes.

In addition to calcium scale, the barium sulfate scale is considered as one of the most frequent and obstinate scales in off shore oil and gas production systems [26]. Because of the extremely low solubility, the barium sulfate scale removal is particularly difficult. The influence of CCQDs to control the barium sulfate formation is shown in **Figure 4c**. It can be seen that CCQDs has the inhibition efficiency of 90 % at 16 mg/L at the temperature of 40 ℃. Different from calcium scales, the scale inhibition efficiency of barium sulfate scales are less affected by temperature. When the dosage of CCQDs is above 20 mg/L, at any temperature from 40℃ to 80℃, the scale inhibition rate reaches 100 %. The possible reason is that barium sulfate has higher thermal stability and its solubility is less affected by temperature.

**Figure 4d** shows the influences of pH on the inhibition efficiency of CCQDs against $CaSO_4$ scales and $BaSO_4$ scales. In this study, the dosage of CCQDs was 40 mg/L and 4 mg/L, respectively. With the increase of pH value, the scale inhibition efficiency for $CaSO_4$ and $BaSO_4$ scales show a low inhibition effect at strong acid conditions. Using 0.5 mol/L sodium hydroxide solution to adjust the solution to alkaline, the CCQDs showed excellent scale inhibition performance. The carboxyl group is terminated on the surface of the carbon quantum dot, which is easier to dissociate and form anions under the alkaline condition and react with scale ions. The scale inhibition efficiency of CCQDs in alkaline is higher than that of in the acidic solution. So the CCQD is more suitable for application in the alkaline solution than in the acid solution.

The scale inhibition effect of CCQDs can be easily observed from the scale inhibition test (**Figure 5**). There are large amount of scales in the beaker and on the beaker wall after heating 10 h without adding CCQDs (**Figure 5a**). In contrast, no scale could be observed after addition of CCQDs with a similar heating time of 10 h (**Figure 5b**).

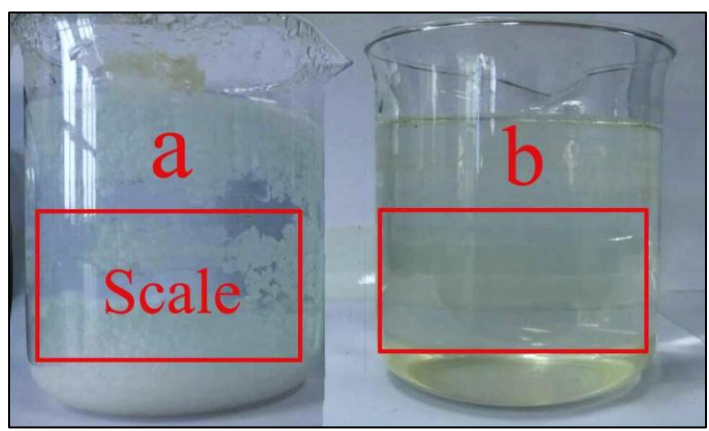

**Figure 5** The photos of CaSO$_4$ solution: (a) free-CCQDs; (b) with CCQDs. Other conditions: T = 80 $^o$C; heating time = 10 h.

In order to eliminate the influence of citric acid to the scale inhibition, a control scale inhibition experiment of citric acid was carried out (**Figure S2**), citric acid had little scale inhibition performance for CaSO$_4$, its scale-inhibition efficiency can only reach 20 % and this number was almost constant with the increasing dosage of citric acid. It was demonstrated that CCQDs synthesized from citric acid has excellent scale-inhibiting ability. As efficient scale inhibitors, polyaspartic acid and polyepoxysuccinic acid have the disadvantages of difficult preparation, strict use conditions and high price. On the contrary, CCQDs are low-toxic, environment friendly, and cheaper to manufacture. It has strong competitiveness in mass production and applications.

To understand the scale inhibition mechanism of carbon quantum dots, XRD (**Figure 6**) and SEM (**Figure 7**) were performed to analyze. As shown in **Figure 6**, the scales formed with free indicators is calcium sulfate dehydrate [27]. After addition of CCQDs in the solution, the calcium sulfate dihydrate was transformed to calcium sulfate hemihydrate. From **Figure 6b**, we can see crystal transformation occurred, there was a peak at $2\theta = 20°$ and the peak at around $2\theta = 11°$ disappeared and the peak at around $2\theta = 15°$ appeared. With the concentration of CCQDs increasing, calcium sulfate hemihydrate was transformed to anhydrous calcium sulfate [27] (**Figure 6c**, **6d**). When the dosage of CCQDs increased from 20 mg to 40 mg, the ratio of peak intensity at $2\theta = 25°$ to peak intensity at $2\theta = 30°$ become weaker. According to this result, it was obviously that the planes at around $2\theta = 25°$ were inhibited. So the addition of CCQDs caused the distortion of calcium sulfate crystal lattice. This result is also confirmed by SEM image of scales (**Figure 7**), the surface of dehydrate calcium sulfate is smooth and the shape is rod-like (**Figure 7a**). After addition of CCQDs, the morphologies of the scales changed greatly. Needle shaped crystallites reduce and disorderly granular scales increase (**Figure 7b**). According to the above results, it can be speculated that lattice distortion happened due to the presence of CCQDs. Morphologies of barium sulfate scales also have changed greatly after addition of CCQDs. In the absence of inhibitor, barium sulfate particles, which are brick-shaped with a width of about 5 μm and the length-width ratio is about 2:1 (**Figure 7c**). When addition of inhibitor (**Figure 7d**), the brick-shaped barium sulfate particles changed to flower-shaped "fleshy plants" with blade length 2-10 μm.

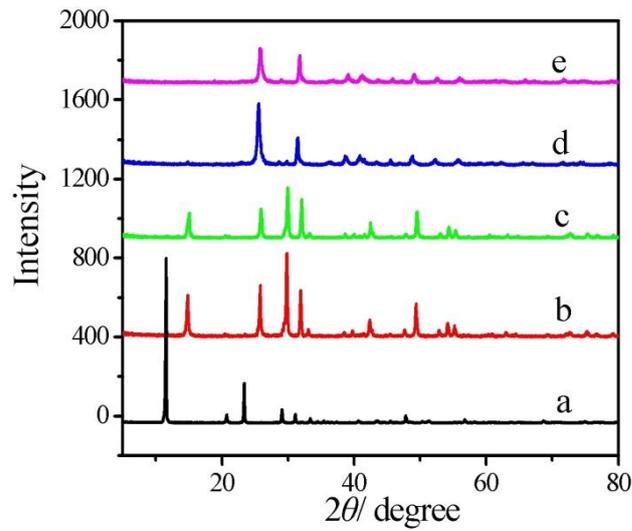

**Figure 6** XRD of calcium sulfate scales with different dosage of CCQDs: (a) 0 mg/L; (b) 5 mg/L; (c) 10 mg/L; (d) 20 mg/L; (e) 40 mg/L.

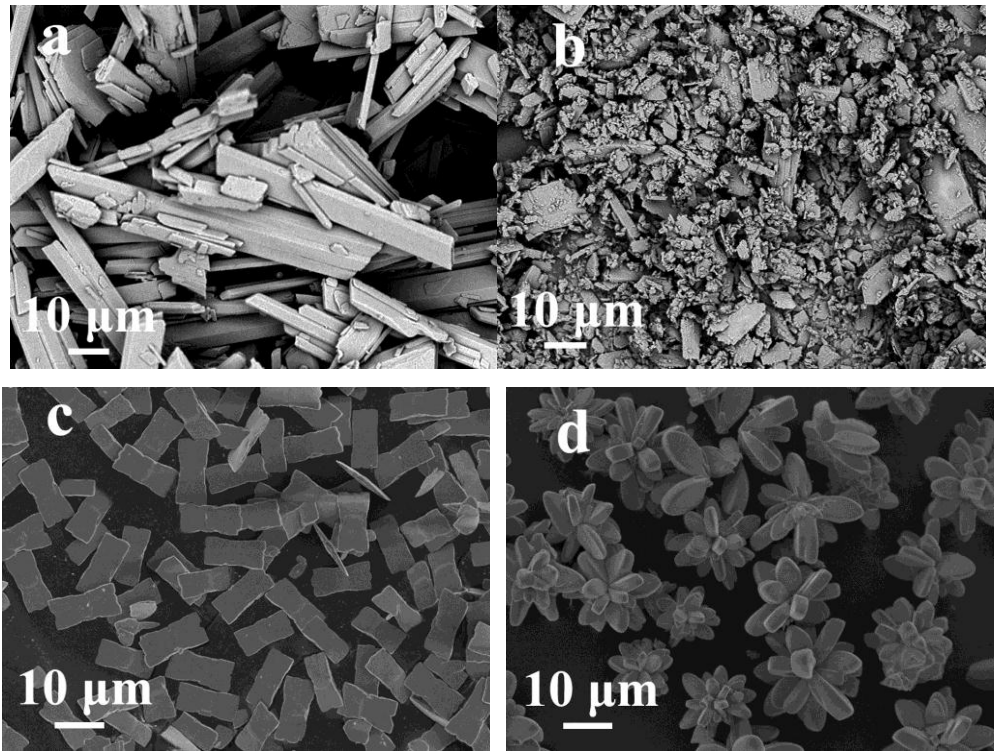

**Figure 7** SEM images of scales: (a) CaSO$_4$, free-CCQDs; (b) CaSO$_4$, with CCQDs 20 mg/L; (c) BaSO$_4$, free- CCQDs; (d) BaSO$_4$, with CCQDs 4 mg/L.

The possible inhibition mechanism according to the above results is depicted in **Figure 8** (calcium sulfate). When CCQDs dissolved in water, negatively charged carboxyl anions chelate with $Ca^{2+}$, which increased the solubility of inorganic salt and hamper the normal growth of inorganic salt crystal, reducing the formation of salt scale. Meanwhile, the CCQDs were adsorbed on the microcrystalline of inorganic salt, which increased the repulsion between particles, hindered their coalescence, and made them in a good dispersion state, so as to prevent or reduce the formation of scale substances. At a smaller size of CCQDs, the Brownian motion is much faster. The rapid movement of carbon quantum dots exacerbated the destruction of calcium sulfate crystals, causing the calcium sulfate grew anomaly, and thus lattice distortion happened. So, calcium sulfate scale was fragmented and disordered.

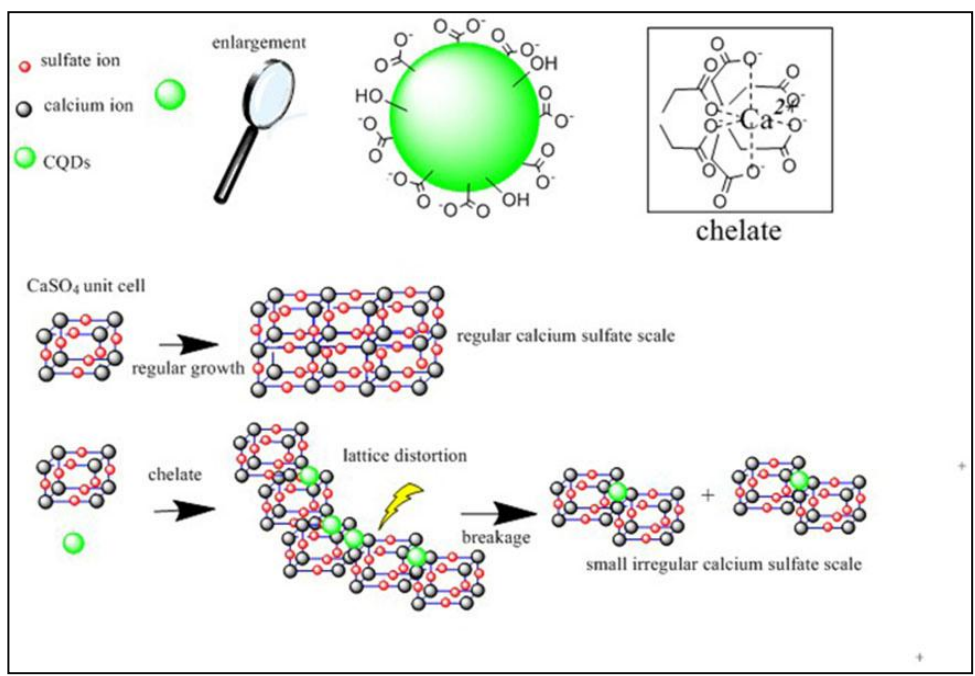

**Figure 8** Possible scale inhibition mechanism of calcium sulfate

## 4. Conclusions

Carbon quantum dots were synthesized by a simple method of thermal decomposition of citric acid. The XPS results reveal that the surface of carbon quantum dots is rich of carboxyl groups. The as-prepared CCQDs exhibit good scale-inhibition properties. The scale inhibition rates of $CaSO_4$ and $BaSO_4$ scales were measured by static scale inhibition method. And the effects of CQDs, constant temperature, pH and other factors on the scale inhibition performance were discussed. As a new type of high efficient scale inhibitor, the CCQDs have many advantages, such as non-toxic, biocompatible and environment friendly. This kind of green scale inhibitor can prevent the formation of scales, improve the efficiency of heat exchange, reduce power or fuel consumption, reduce the sewage water treatment and improve water use efficiency. It will bring great influence on many application fields, such as water treatment, desalination of sea water, metal smelting, petrochemical industry etc.

**Acknowledgements**

Authors thank the financial support of National Nature Science Foundation (No. 21376178), Tianjin Innovation and Development of Regional Marine Economy Demonstration project (cxsf2014-26), and the Youth Innovation Foundation of Tianjin University of Science & Technology (No. 2015LG15). Authors thank Dr. Xiaobin Lian (Quanzhou Normal University) for helping collection and analysis of the XPS experimental data. Authors thank Prof. Bin Ren (Xiamen University) for his English correction and good advice.

**Graphical Abstract**

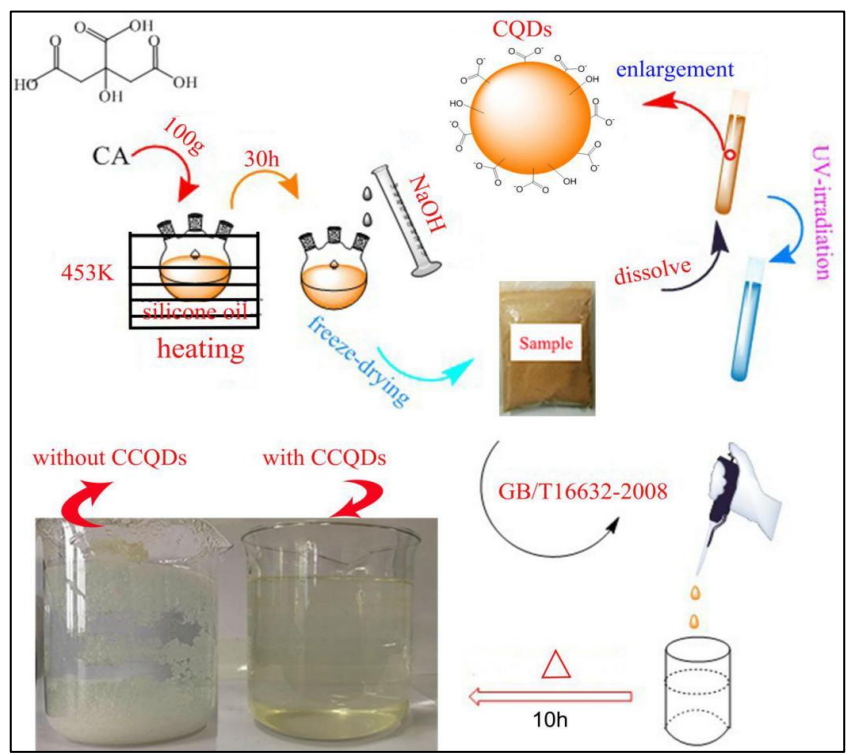

# Supporting Information

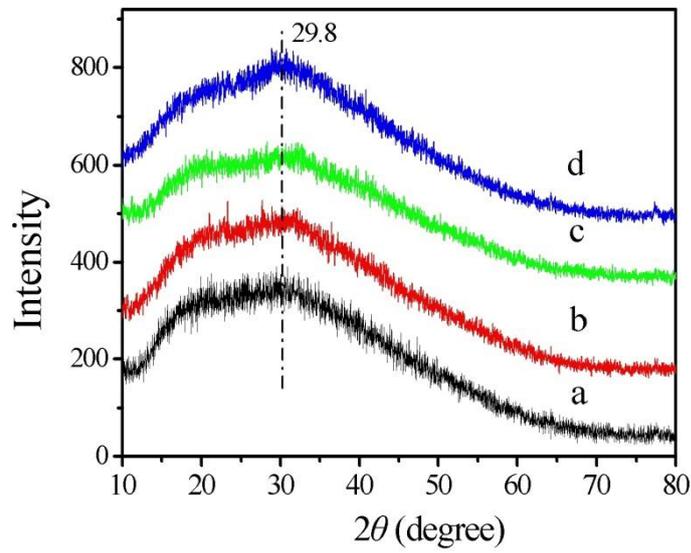

**Fig. S1.** The X-ray diffraction of CCQDs obtained by different pyrolysis time: (a) 10 h, (b) 20 h, (c) 30 h, and (d) 40 h.

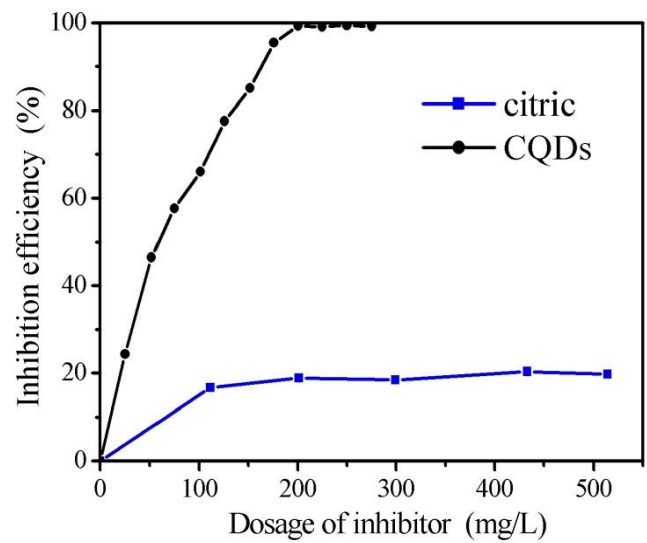

**Fig. S2.** The scale inhibition efficiency of CCQDs and citric acid for calcium sulfate.

Table S1. The atomic percent (at. %) of CCQDs from XPS (H was not considered).

|  | 10-h CCQDs | 20-h CCQDs | 30-h CCQDs | 40-h CCQDs |
|---|---|---|---|---|
| C (at. %) | 60.6 | 62.5 | 66.1 | 66.6 |
| O (at. %) | 39.4 | 37.5 | 33.9 | 33.4 |